\documentclass[aps,prd,final,nofootinbib,superscriptaddress]{revtex4}
\usepackage{amsmath,amsfonts,amsthm,amssymb,graphicx,epsfig,bm,mathrsfs}
\usepackage{bbold}
\usepackage{color}
\usepackage{ytableau}
\usepackage{parskip}
\usepackage{fancyhdr}
\usepackage[utf8]{inputenc}
\usepackage{hyperref}
\usepackage{ulem}

\theoremstyle{plain}
\theoremstyle{definition}
\def\be{\begin{equation}}
\def\ee{\end{equation}}
\def\ba{\begin{eqnarray}}
\def\ea{\end{eqnarray}}

\def\bdm{\begin{displaymath}}
\def\edm{\end{displaymath}}

\def\bq{\begin{quote}}
\def\eq{\end{quote}}

\def\d{{\rm d}}
\def\del{\partial}
\def\ltap{\ \raise.3ex\hbox{$<$\kern-.75em\lower1ex\hbox{$\sim$}}\ }
\def\gtap{\ \raise.3ex\hbox{$>$\kern-.75em\lower1ex\hbox{$\sim$}}\ }
\def\gl{\ \raise.5ex\hbox{$>$}\kern-.8em\lower.5ex\hbox{$<$}\ }
\def\roughly#1{\raise.3ex\hbox{$#1$\kern-.75em\lower1ex\hbox{$\sim$}}}

 at 11truept

\def \plethysm {\sym^m(\bigwedge^p) \otimes \sym^{m-1}(\sym^2)}

\def\d2{{(\hat D \pi)^2 }}

\def \E {{\mathcal E}}
\def \A {{\mathcal A}}

\def \ap {A}
\def \bp {B}

\def \F {{\mathcal F}}

\def \gl {D}

\newcommand{\sym}{\operatorname{Sym}}

\newcommand{\beq}{\begin{equation}}
\newcommand{\eeq}{\end{equation}}
\newcommand{\bea}{\begin{eqnarray}}
\newcommand{\eea}{\end{eqnarray}}
\newcommand{\beqa}{\begin{eqnarray}}
\newcommand{\eeqa}{\end{eqnarray}}

\def \1 {\pi_{i_1}}
\def \2 {\pi_{i_2}}
\def \3 {\pi_{i_3}}
\def \4 {\pi_{i_4}}
\def \5 {\pi_{i_5}}
\def \6 {\pi_{i_6}}
\def \7 {\pi_{i_7}}

\begin{document}
\hspace{4.5in} \mbox{YITP-17-35, IPMU17-0052}\\
%\vspace{-1.03cm} % Preprint number

\title{Classifying Galileon $p$-form theories}
\author{C\'edric Deffayet}
\affiliation{UPMC-CNRS, UMR7095, Institut d'Astrophysique de Paris, GReCO, 98bis boulevard Arago, F-75014 Paris, France}
\affiliation{IHES, Le Bois-Marie, 35 route de Chartres, F-91440 Bures-sur-Yvette, France}
\author{Sebastian Garcia-Saenz}
\affiliation{UPMC-CNRS, UMR7095, Institut d'Astrophysique de Paris, GReCO, 98bis boulevard Arago, F-75014 Paris, France}
\author{Shinji Mukohyama}
\affiliation{Center for Gravitational Physics, Yukawa Institute for Theoretical Physics, Kyoto University, 606-8502, Kyoto, Japan}
\affiliation{Kavli Institute for the Physics and Mathematics of the Universe (WPI), The University of Tokyo Institutes for Advanced Study, The University of Tokyo, Kashiwa, Chiba 277-8583, Japan}
\author{Vishagan Sivanesan}
\affiliation{UPMC-CNRS, UMR7095, Institut d'Astrophysique de Paris, GReCO, 98bis boulevard Arago, F-75014 Paris, France}

\ytableausetup 
{mathmode, boxsize=0.5cm, centertableaux}

\begin{abstract}
We provide a complete classification of all abelian gauge invariant $p$-form theories with equations of motion depending only on the second derivative of the field---the $p$-form analogues of the Galileon scalar field theory. We construct explicitly the nontrivial actions that exist for spacetime dimension $D\leq11$, but our methods are general enough and can be extended to arbitrary $D$. We uncover in particular a new $4$-form Galileon cubic theory in $D\geq8$ dimensions. As a by-product we give a simple proof of the fact that the equations of motion depend on the $p$-form gauge fields only through their field strengths, and show this explicitly for the recently discovered $3$-form Galileon quartic theory.
\end{abstract}

\maketitle

\section{Introduction}

\noindent
Much effort has been put in recent years to the study of scalar field theories with nonlinear equations of motion of at most second order. The construction and complete classification of these models in arbitrary dimensions was achieved in \cite{Deffayet:2011gz}, although they have actually been known, at least in four or less dimensions, since much before \cite{Horndeski:1974wa,Fairlie:1992nb,Fairlie:1992he,Fairlie:1991qe} (see e.g.\ \cite{Deffayet:2013lga} for a review of their formal aspects). This renewed interest is mostly due to the works on the so-called Galileons \cite{Nicolis:2008in}: scalar fields in flat spacetime with equations of motion that depend only on the second derivative of the field. Such fields enjoy a generalized shift symmetry of the form $\pi(x)\to\pi(x)+c+b_ax^a$ (where $c$ and $b_a$ are constants) that has a number of interesting theoretical consequences \cite{Nicolis:2010se,Goon:2011qf,Hui:2012qt,Goon:2012mu,Goon:2012dy,deRham:2013hsa,Creminelli:2013ygt,deRham:2014lqa,Klein:2015iud,Goon:2016ihr}, which, together with their rich phenomenology in the context of modified gravity, has spurred an intense search for generalizations of the original Galileon theory. The $p$-form analogues of the scalar Galileon were first studied in \cite{Deffayet:2010zh}, a work which also provided a framework allowing for extending the original Galileon to multiple scalars (to be considered as $0$-forms). Such multi-scalar extensions, with or without an internal symmetry group, have been examined in depth \cite{Padilla:2010de,Padilla:2010tj,Padilla:2010ir,Hinterbichler:2010xn,Trodden:2011xh,Padilla:2012dx,deFromont:2013iwa,Garcia-Saenz:2013gya,Sivanesan:2013tba,Allys:2016hfl}, and modifications of the Galileon shift symmetry have also been considered \cite{Hinterbichler:2015pqa,Noller:2015rea,Novotny:2016jkh}. The generalization to tensor fields of generic symmetry type has been recently tackled in \cite{Chatzistavrakidis:2016dnj}.

The method developed in \cite{Deffayet:2010zh} allows to construct nontrivial theories for $p$-form fields via actions that pattern the scalar Galileon terms through the schematic replacement $\partial_a\pi\to\F_{a_1\cdots a_{p+1}}$, where $\bm{\F} = d\bm{\A}$ is the abelian field strength. For instance the simplest case is a single-field $2$-form quartic theory given by
\be \label{eq:s24intro}
S_{2,4}=\frac{1}{12}\int d^7x\,\epsilon^{a_1\cdots a_7}\epsilon^{b_1\cdots b_7}\partial_{a_1}\F_{b_2b_3b_4}\partial_{b_1}\F_{a_2a_3a_4}\F_{b_5b_6b_7}\F_{a_5a_6a_7}\,,
\ee
where the notation $S_{p,m}$ indicates the rank $p$ of the form field and the order $m$ of the interaction, and we have written the action for the minimum allowed value of the spacetime dimension $D$, i.e.\ $D=7$. Note that above and henceforth, the tensor $\epsilon$ is the totally antisymmetric Levi-Civita tensor defined as usual in flat $D$ dimensions by
\ba \label{DEFLC}
\epsilon^{a_{\vphantom{()}1} a_{\vphantom{()}2} \ldots
a_{\vphantom{()}D}} \equiv 
\delta^{[a_{\vphantom{()}1}}_0 \delta^{a_{\vphantom{()}2}}_1
\ldots \delta^{a_{\vphantom{()}D}]}_{D-1}\,.
\ea
This is an important difference between the scalar and $p$-form cases: whereas interactions among Galileon scalars exist for any $D\geq2$, Galileon $p$-forms are strongly constrained by the dimension. The above construction technique is not exhaustive, as it is restricted---when a single $p$-form is considered---to even $p$ values and to interactions of even order $m$, quartic and higher. In fact, it was later shown that nontrivial actions don't exist at all for $p=1$, i.e.\ for gauge invariant vector Galileons \cite{Deffayet:2013tca},\footnote{Nontrivial vector theories do exist if one relaxes the assumption of gauge invariance \cite{Tasinato:2014eka,Heisenberg:2014rta,Hull:2014bga,Hull:2015uwa,Khosravi:2014mua,Allys:2015sht,Jimenez:2016isa,Heisenberg:2016eld,Allys:2016jaq,Kimura:2016rzw,Allys:2016kbq,Jimenez:2016upj}.} but fortunately this no-go result doesn't extend to higher odd $p$ values, as a $3$-form quartic vertex has been recently found in \cite{Deffayet:2016von}, which we review below.

The existence of this nonlinear $3$-form theory was established through a general group theoretic argument whose consequences were not fully explored in ref.\ \cite{Deffayet:2016von}. It is the main purpose of the present work to fulfill this task by providing a complete classification and explicit construction of all Galileon $p$-form theories that exist for $D\leq11$. This value of $D$ was chosen having in mind upper bounds that arise in the contexts of string theory and supergravity, but also just for the sake of being concrete as we will argue that our methods can be easily extended to higher dimensions. Our results include, in particular, a previously unknown cubic $4$-form Galileon theory that exists in $D\geq8$. Another by-product of the results of \cite{Deffayet:2016von} that we uncover is a simple proof of the well-known fact that the gauge invariance of any $p$-form theory implies that the equation of motion can be written in terms of the field strength $\bm{\F}$ alone \cite{Henneaux:1997ha,Henneaux:1999ma}. We show this explicitly for the actions in our classification by noting that they are all of the ``Born--Infeld'' type, that is they are themselves functionals of $\bm{\F}$ only.

The contents of the rest of the paper are as follows. As a follow-up to the introduction, we briefly review in section \ref{sec:3formexample} the construction technique of ref.\ \cite{Deffayet:2010zh} and show why it fails for odd $p$-forms. We then present the $3$-form theory recently found in \cite{Deffayet:2016von} and discuss some of its properties. In section \ref{sec:classification} we recall the main theorems established in the latter reference, and we apply them to derive the set of candidate tensors (more specifically, their symmetry type) from which nontrivial Galileon field equations can be found. The explicit construction of these equations and the corresponding actions are given in section \ref{sec:construction}. We conclude with some final comments in section \ref{sec:discussion}.

{\it Conventions:} We use the ``mostly-plus'' metric signature $(-,+,\ldots,+)$ and spacetime indices are denoted by latin letters. Occasionally we will use the notation $a[n]$ to represent an antisymmetrized string of $n$ indices, e.g.\ $\A_{a[p]}=\A_{[a_1\cdots a_p]}$, where (anti)symmetrization of indices is defined with unit weight. Such an antisymmetrized string of indices will also sometimes just be denoted by a capital latin letter without any reference to its length (which can be inferred from the context), e.g.\ $A_1 \equiv \{a_1,\cdots, a_p\}$. The field strength of a $p$-form $\A_{a_1\cdots a_p}$ is given by $\F_{a_1\cdots a_{p+1}}\equiv (p+1)\partial_{[a_1}\A_{a_2\cdots a_{p+1}]}$.

\section{A Galileon 3-form theory} \label{sec:3formexample}

\noindent
We begin by recalling the method of \cite{Deffayet:2010zh} for the construction of interactions of a single $p$-form gauge field $\bm{\A}\,$, and show why it doesn't work when $p$ is odd. Consider for concreteness the quartic scalar, or $0$-form, Galileon vertex in $D=3$ (the minimal allowed value for which the action is nontrivial):\footnote{The coefficient in front the action is of course irrelevant as we are studying interaction vertices independently. We choose to fix them so as to get a unit coefficient in the corresponding equation of motion, as we will make explicit in section \ref{sec:construction}.}
\be
S_{0,4}=\frac{1}{4}\int d^3x\,\epsilon^{a_1a_2a_3}\epsilon^{b_1b_2b_3}(\partial_{a_1}\partial_{b_2}\pi)(\partial_{b_1}\partial_{a_2}\pi)(\partial_{a_3}\pi)(\partial_{b_3}\pi)\,.
\ee
One can then obtain a quartic vertex for any form field of even degree $p$ by mimicking the above structure with the field strength $\F_{a[p+1]}$ in place of $\partial_a\pi$. For example for $p=2$ one gets the action $S_{2,4}$ that we displayed above in (\ref{eq:s24intro}). As mentioned before, this of course imposes a lower bound on the spacetime dimension $D$. The gauge invariance of $S_{2,4}$ is manifest, while the fact that it leads to only two-derivative terms in the equation of motion follows from the Bianchi identity $\partial_{[a_1}\F_{a_2\cdots a_{p+2}]}=0$ (which for a scalar $\pi$, i.e.\ a $0-$form, is simply the ``Schwarz theorem" stating that $\partial_{[a_1}\partial_{a_2]} \pi$ vanishes). It is also easy to see that this prescription fails when $p$ is odd. For instance if $p=3$ one would have
\be \label{eq:s34pintro}
S_{3,4}'=\frac{9}{64}\int d^9x\,\epsilon^{a_1\cdots a_9}\epsilon^{b_1\cdots b_9}\partial_{a_1}\F_{b_2b_3b_4b_5}\partial_{b_1}\F_{a_2a_3a_4a_5}\F_{b_6b_7b_8b_9}\F_{a_6a_7a_8a_9}\,,
\ee
which in fact vanishes identically, since an integration by parts and a relabeling of indices produces $S_{3,4}'=-S_{3,4}'$ \cite{Deffayet:2010zh}.

A more general procedure was developed in \cite{Deffayet:2016von}, which we review and apply in section \ref{sec:classification}, to identify candidate Galileon $p$-form interaction vertices without any a priori assumptions regarding the contraction of indices in the Lagrangian. A particularly interesting result was the identification and explicit construction of a quartic $3$-form theory, which can be written as
\be \label{eq:s34intro}
S_{3,4}=\frac{1}{4}\int d^9x\,\epsilon^{a_1\cdots a_9}\epsilon^{b_1\cdots b_9}\F_{a_1a_2a_3a_4}\F_{b_1b_2b_3b_4}\partial_{a_5}\partial_{b_6}\A_{b_7b_8a_9}\partial_{b_5}\partial_{a_6}\A_{a_7a_8b_9}\,.
\ee
We are again displaying the action for the minimum allowed value of the dimension; the higher dimensional version is obtained simply by contracting the additional indices in the Levi-Civita tensors. Absence of higher derivative terms in the field equation that derives from (\ref{eq:s34intro}) follows from the same reasons as above, while invariance under the abelian gauge symmetry $\delta{\bm\A}=d{\bm\Lambda}$ is also immediate due to the antisymmetry of the ${\bm\epsilon}$ tensors. In fact, this action is a functional of the field strength only---it is of the Born--Infeld type \cite{Henneaux:1997ha}\footnote{We thank N.\ Boulanger for insightful comments on this point.}---which can be seen by noting that
\be
\epsilon^{a_1\cdots}\epsilon^{b_1\cdots}\partial_{a_1}\F_{b_1b_2b_3a_2}=\frac{4}{3}\,\epsilon^{a_1\cdots}\epsilon^{b_1\cdots}\partial_{a_1}\partial_{b_1}\A_{b_2b_3a_2}\,.
\ee
Thus we can rewrite (\ref{eq:s34intro}) as
\be \label{eq:s34intro2}
S_{3,4}=\frac{9}{64}\int d^9x\,\epsilon^{a_1\cdots a_9}\epsilon^{b_1\cdots b_9}\F_{a_1a_2a_3a_4}\F_{b_1b_2b_3b_4}\partial_{a_5}\F_{b_6b_7b_8a_9}\partial_{b_5}\F_{a_6a_7a_8b_9}\,.
\ee
We will come back to this point regarding manifest gauge invariance in section \ref{sec:construction} where we will rederive, via a simple argument, the result that the equation of motion of a $p$-form gauge theory must be expressible in terms of ${\bm\F}$.

The difference between the vanishing action (\ref{eq:s34pintro}) and the nontrivial one in (\ref{eq:s34intro2}) is formally very simple---one simply contracts indices in a different way. A natural question is then whether {\it other} contractions could yield dynamically independent actions. It is this the aspect for which the group theoretic results of ref.\ \cite{Deffayet:2016von}, which we next review and apply, are especially useful. Indeed, they will allow us to fully classify and construct all the Galileon $p$-form actions that exist up to $D=11$ and in particular to show how seemingly different theories can in fact be related. We will also illustrate this in an explicit way when we go through the bottom-up construction of the interaction vertices in section \ref{sec:construction}.

\section{Classifying possible {\it p}-form Galileon interactions} \label{sec:classification}

\noindent
In Ref.~\cite{Deffayet:2016von}, necessary conditions for the existence of a non trivial Galileon $p$-form theory have been derived. Namely, one first defines the field equation operator $\bm{\E}$ (of components denoted by $\E^{\ap}$, such that the field equation for the $p$-form $\bm{\A}$ reads $\E^{\ap} = 0$), and furthermore denoted by $\bm{\E^m}$ the tensor obtained by differentiating $\bm{\E}$ with respect to the second derivatives of the $p$-form $m-1$ times; i.e., such that one has by definition $\bm{\E^1}\equiv \bm{\E}$ and in full generality $\bm{\E^m}$ is the rank $(pm + 2(m-1))$ contravariant tensor whose components are given by 
\be \label{defEm}
\left(\bm{\E^m}\right)^{\ap \bp_1 c_1d_1 \dots \bp_{m-1} c_{m-1}d_{m-1}}
%= \E^{\ap|\bp_1,c_1d_1|\dots |\bp_{m-1},c_{m-1}d_{m-1}} 
\equiv \frac{\del^{m-1} \E^{\ap}}{\del \A_{\bp_1,c_1d_1} \dots \del \A_{\bp_{m-1},c_{m-1}d_{m-1}}}\,,
\ee
where here and henceforth the notation ``$\A_{\bp,cd}$" denotes $\partial_c \partial_d \A_{B}$. It was then shown that in order for such a theory to (i) derive from an action, (ii) have field equations which are gauge invariant (in the abelian sense) and which (iii) {\it only} contain second derivatives of the $p$-form, the tensors $\bm{\E^m}$ must (for any $m$) have certain nontrivial symmetries. In particular, this tensor must (for any values of $m$) belong to the ``plethysm"  $\plethysm$, where $\sym^m(\bigwedge^p)$ denotes the linear span of tensors made by the $m$ times symmetric tensor products of $p$-forms, while $\sym^{m-1}(\sym^2)$ denotes the linear span of tensors made by the $m-1$ times symmetric tensor products of symmetric 2-tensors (see Ref.~\cite{Deffayet:2016von} for more details). It was further shown that, in order for a theory to obey properties (i), (ii) and (iii) above, the operators $\bm{\E^m}$ must not only belong to the aforementioned plethysm, but also correspond to irreducible representations of the symmetric group (acting on the free indices of these operators) associated to Young diagrams with no more than two columns (more properly, as explained in \cite{Deffayet:2016von}, the operators $\bm{\E^m}$ must belong to the tensor symmetry classes associated with these diagrams). A general formula giving the multiplicity $M_{(mp + 2(m-1)-a,a)^t}$ of the irreducible corresponding to the Young diagram $(mp + 2(m-1)-a,a)^t$ inside the plethysm $\plethysm$ was then derived, where $(b,a)^t$ denotes here and in the following a Young diagram with two columns, the first with $b$ boxes and the second with $a \leq b$ boxes. The multiplicity $M_{(mp + 2(m-1)-a,a)^t}$ was found to be given by 
\be \label{finalrescount}
M_{(mp + 2(m-1)-a,a)^t} =
\begin{cases}
N^p_{a-m+1,m} - N^p_{a-m,m} & \text{if $p$ is even}\,,\\
N^{p,distinct}_{a-m+1,m} - N^{p,distinct}_{a-m,m} & \text{ if $p$ is odd}\,,
\end{cases}
\ee
where for $r \geq 0$, $N^p_{r,s}$ is the number of (unordered) partitions of $r$ into $s$ non-negative integers within $\{0,\cdots,p\}$ with repetitions allowed and $N^{p,distinct}_{r,s}$ is the number of (unordered) partitions of $r$ into $s$ distinct non-negative integers within $\{0,\cdots,p\}$; and we have $N^p_{r,s} =0 $ and $N^{p,distinct}_{r,s} =0$ for negative $r$. In particular the multiplicity vanishes for $a-m+1<0$ and thus we only need to consider $a$ in the range $m-1\leq a\leq [mp+2(m-1)]/2$. Moreover, for odd $p$, the multiplicity vanishes if $a-m+1<\sum_{i=0}^{m-1}i$. Thus, for odd $p$, we only need to consider $a$ in the range $(m-1)(m+2)/2\leq a \leq  [mp+2(m-1)]/2$. For a given diagram, the corresponding multiplicity given above gives the number of isomorphic irreducible representations, each indexed by a different standard Young tableau built from the corresponding Young diagram, which enters in the decomposition of the plethysm $\plethysm$ of interest here. 

If, for a given $p$, there exists a non trivial $p$-form theory (i.e.\ a theory verifying properties (i), (ii) and (iii) above) then this theory must have at least a non vanishing tensor $\bm{\E^2}$ (this because, with our conventions $\bm{\E^1}$ are just the field equations, and if non trivial field equations exists at all, they must be at least linear in the second derivatives, meaning here that at least  $\bm{\E^2}$ must be non vanishing). Moreover, a non trivial Galileon $p$-form theory must have at least a non vanishing $\bm{\E^3}$, otherwise the field equations would be at most linear in second derivatives. Hence, we can restrict ourselves to 
\be
m \geq 2 \label{mnontrivp}
\ee
if we look for a non trivial $p$-form theory, and to 
\be
m \geq 3 \label{mnontrivGalp}
\ee
if we look for a non trivial Galileon p-form theory, and we have to check in each cases that there are existing (i.e.\ with non vanishing multiplicities) representations corresponding to the tensors $\bm{\E^m}$. For this to happen, we have further shown in Ref.\ \cite{Deffayet:2016von} that one must have for odd $p$ 
\be \label{eq:mupperbound}
m \leq p+1 \qquad\mbox{(for odd $p$)}\,.
\ee
This last condition shows in particular that no nontrivial vector Galileon exists, a result first obtained in Ref.\ \cite{Deffayet:2013tca} in a different way. Moreover, for a given space-time dimension $D$, we must have enough different indices values at hand to ``fill-up" all the cases of a tensor belonging to the symmetry class indexed by a given standard tableau built from a given Young diagram $(mp + 2(m-1)-a,a)^t$. This means that one must have simultaneously (this condition coming from the antisymmetry associated with the cases belonging to a given column of the Young diagram)
\be
\begin{aligned}
0\leq mp +2(m-1)-a &\leq D\,, \\
0 \leq a &\leq D\,.
\end{aligned}
\ee
The above two constraints translate into the bound
\be
p \leq 2 \;\frac{(D+1-m)}{m}\,, \label{eqn:p_max}
\ee
or equivalently,
\be
m \leq 2 \; \frac{D+1}{p+2}\,. \label{eqn:m_max}
\ee
Using then (\ref{mnontrivp}) and (\ref{mnontrivGalp}) we get that for a given space-time dimension $D$ a non trivial p-form theory can only exists (using the above bound with $m=2$) if 
\be p \leq (D-1)
\ee 
(which is more restrictive than the ``naive" bound $p \leq D$ and accounts for the presence of derivatives in the field equations), and a nontrivial Galileon $p$-form can only exist (using the above bound with $m=3$) if 
\be p \leq 2 (D-2)/3\,,
\ee
while we also see that in order for the field equations to depend non trivially on higher powers of the second derivatives of the $p$-form, then the admissible values of $p$ become more restricted. As explained in the introduction (see also our final comments in section \ref{sec:discussion}), we will further restrict in this work our attention to spacetime dimensions up to $D=11$, while so far all our bounds apply in any dimension. We then apply formula \eqref{finalrescount} for all values of $p$ and $m$ (restricted here to $m \geq 2$ for the reasons given above) in order to determine which diagrams $(mp + 2(m-1)-a,a)^t$ have a nonzero multiplicity. The results are shown in table \ref{tab:youngtable-all} below.
\begin{table}[h]
\centering
\begin{tabular}{|c|c|c|c|c|c|}
\hline
     & $m=2$                                    & $m=3$                                     & $m=4$                          & $m=5$                   & $m=6$                   \\ \hline
    $p=2$  & ${{(3,3)}^t}, {{(5,1)}^t}$                     & ${{(6,4)}^t}, {{(8,2)}^t}$                      & ${{(7,7)}^t}, {{(9,5)}^t}, {{(11,3)}^t}$ & ${{(10,8)}^t}{, {{(12,6)}^t}}$  & ${{(11,11)}^t}{, {{(13,9)}^t}}$    \\
& & & &  { ${{(14,4)}^t}$} & { ${{(15,7)}^t}, {{(17,5)}^t}$}\\ \hline
$p=3$  & ${{(4,4)}^t}, {{(6,2)^t}}$                     & ${{(8,5)}^t}$                               & ${{(9,9)}^t}$                    &                       &                       \\ \hline
$p=4$  & ${{(5,5)}^t}, {{(7,3)}^t}, {{(9,1)}^t}$            & ${{(8,8)}^t}, {{(10,6)}^t}, {{(11,5)}^t}$ & ${{(11,11)}^t}{, {{(13,9)}^t}\times 2}$                  &                       &                       \\ 
& & { ${{(12,4)}^t}, {{(14,2)}^t}$} & { ${{(15,7)}^t}\times 2, {{(16,6)}^t}$} & & \\
& & &{ ${{(17,5)}^t}, {{(19,3)}^t}$} & & \\ \hline
$p=5$  & ${{(6,6)}^t}, {{(8,4)^t}}, {{(10,2)^t}}$           & ${{(11,8)^t}}{ ,{{(12,7)^t}}, {{(14,5)^t}}}$                    &                              &                       &                       \\ \hline
$p=6$  & ${{(7,7)}^t}, {{(9,5)^t}}, {{(11,3)^t}}$ & { ${{(12,10)^t}}, {{(14,8)^t}}\times 2$}                            &                              &                       &                       \\
& { ${{(13,1)^t}}$} & { ${{(15,7)^t}}, {{(16,6)^t}}, {{(17,5)^t}}$}& & &\\
& &  { ${{(18,4)^t}}, {{(20,2)^t}}$}  & & &\\ \hline
$p=7$  & ${{(8,8)}^t}, {{(10,6)^t}}{, {{(12,4)^t}}}$          & \multicolumn{1}{l|}{}                   & \multicolumn{1}{l|}{}        & \multicolumn{1}{l|}{} & \multicolumn{1}{l|}{} \\ 
& { ${{(14,2)^t}}$}& & & &\\ \hline
$p=8$  & ${{(9,9)}^t}, {{(11, 7)^t}}{, {{(13,5)^t}}}$                  & \multicolumn{1}{l|}{}                   & \multicolumn{1}{l|}{}        & \multicolumn{1}{l|}{} & \multicolumn{1}{l|}{} \\ 
& { ${{(15,3)^t}}, {{(17,1)^t}}$}& & & &\\ \hline
$p=9$  & ${{(10,10)}^t}{, {{(12,8)^t}}, {{(14,6)^t}}}$                  & \multicolumn{1}{l|}{}                   & \multicolumn{1}{l|}{}        & \multicolumn{1}{l|}{} & \multicolumn{1}{l|}{} \\ 
& { ${{(16,4)^t}}, {{(18,2)^t}}$}& & & &\\ \hline
$p=10$ & ${{(11,11)}^t}{, {{(13,9)^t}}, {{(15,7)^t}}}$                            &                                         &                              &                       &                       \\ 
& { ${{(17,5)^t}}, {{(19,3)^t}}, {{(21,1)^t}}$}& & & &\\ \hline
\end{tabular}
\caption{Young diagrams which respect the bounds (\ref{eqn:p_max}) and (\ref{eqn:m_max}) with $D\leq 11$ and for which the associated representations exist in the plethysm $\plethysm$, as inferred from the formula \eqref{finalrescount}. The representations ${{(13,9)^t}}$ and ${{(15,7)^t}}$ for $p=4$, $m=4$ and the representation ${{(14,8)^t}}$ for $p=6$, $m=3$ have multiplicity $2$. All other representations appearing here have multiplicity $1$.}
\label{tab:youngtable-all}
\end{table}

The multiplicities shown in table \ref{tab:youngtable-all} satisfy the following consistency relation that follows from the formula \eqref{finalrescount}:
\begin{equation}
\sum_{a=0}^{a_{\mathrm{max}}(m,p)} M_{(mp + 2(m-1)-a,a)^t} = N^p_{a_{max}(m,p)-m+1,m}\,,
\end{equation}
where $a_{\mathrm{max}}(m,p)=[mp+2(m-1)]/2$ for even $mp$ and $a_{\mathrm{max}}(m,p)=[mp-1+2(m-1)]/2$ for odd $mp$. However, some of the Young diagrams in table \ref{tab:youngtable-all} have more than $11$ boxes in the first column and they are not of our interest as we have assumed $D\leq 11$. Excluding those with more than $11$ boxes in the first column, we obtain table \ref{tab:youngtable} below, which is fully consistent with the generic bounds given above.

\begin{table}[h]
\centering
\begin{tabular}{|c|c|c|c|c|c|}
\hline
     & $m=2$                                    & $m=3$                                     & $m=4$                          & $m=5$                   & $m=6$                   \\ \hline
    $p=2$  & ${\mathbf{(3,3)}^t}, {\mathbf{(5,1)}^t}$                     & ${\mathbf{(6,4)}^t}, {\mathbf{(8,2)}^t}$                      & ${\mathbf{(7,7)}^t}, {\mathbf{(9,5)}^t}, {\mathbf{(11,3)}^t}$ & ${\mathbf{(10,8)}^t}$  & ${\mathbf{(11,11)}^t}$           \\ \hline
$p=3$  & ${\mathbf{(4,4)}^t}, {{(6,2)^t}}$                     & ${\mathbf{(8,5)}^t}$                               & ${\mathbf{(9,9)}^t}$                    &                       &                       \\ \hline
$p=4$  & ${\mathbf{(5,5)}^t}, {\mathbf{(7,3)}^t}, {\mathbf{(9,1)}^t}$            & ${\mathbf{(8,8)}^t}, {\mathbf{(10,6)}^t}, {{(11,5)}^t}$ & ${\mathbf{(11,11)}^t}$                  &                       &                       \\ \hline
$p=5$  & ${\mathbf{(6,6)}^t}, {{(8,4)^t}}, {{(10,2)^t}}$           & ${{(11,8)^t}}$                    &                              &                       &                       \\ \hline
$p=6$  & ${\mathbf{(7,7)}^t}, {{(9,5)^t}}, {{(11,3)^t}}$ &                             &                              &                       &                       \\ \hline
$p=7$  & ${\mathbf{(8,8)}^t}, {{(10,6)^t}}$          & \multicolumn{1}{l|}{}                   & \multicolumn{1}{l|}{}        & \multicolumn{1}{l|}{} & \multicolumn{1}{l|}{} \\ \hline
$p=8$  & ${\mathbf{(9,9)}^t}, {{(11, 7)^t}}$                  & \multicolumn{1}{l|}{}                   & \multicolumn{1}{l|}{}        & \multicolumn{1}{l|}{} & \multicolumn{1}{l|}{} \\ \hline
$p=9$  & ${\mathbf{(10,10)}^t}$                  & \multicolumn{1}{l|}{}                   & \multicolumn{1}{l|}{}        & \multicolumn{1}{l|}{} & \multicolumn{1}{l|}{} \\ \hline
$p=10$ & ${\mathbf{(11,11)}^t}$                            &                                         &                              &                       &                       \\ \hline
\end{tabular}
\caption{Young diagrams for which the number of boxes in each column is equal to or less than $11$ and the associated representation exists in the plethysm $\plethysm$, as inferred from formula \eqref{finalrescount}. All the representations appearing here have multiplicity $1$. The difference between the diagrams indicated in boldface and plain font is explained in section \ref{sec:construction}. Those in boldface either have the same number of boxes in the two columns or satisfy the condition (\ref{eqn:redundant}). While each of the former defines a distinct theory, the latter are shown to be redundant. Those in plain font do not lead to nontrivial equations of motion that depend only on the second derivative of the $p$-form.}
\label{tab:youngtable}
\end{table}

Once we know that for a given $p$, $m$ and $D$ there exists a non-void tensor symmetry class with the required symmetry (which is indicated in the above table by a case containing a Young diagram as opposed to a void case)  the next step is to show that a corresponding $\bm{\E^m}$ can indeed be built from the tensors at hand, namely the metric, the Levi-Civita tensor, the $p$-form and its derivatives. This is what we do in the next section. Note however for now that if a given $\bm{\E^m}$ with the required symmetry exists, it can a priori be both integrated and differentiated with respect to the second derivatives of the $p$-form, yielding respectively nontrivial tensors $\bm{\E^{m-1}}$ and $\bm{\E^{m+1}}$ having the same symmetries (once the differences in the number of spacetime indices are taken into account) and correspond to the same theory. This means that for every diagram present in the above table in some case with a given value of $p$ and $m$ there should also be generically a diagram indicated for the same value of $p$ and $m-1$ or $m+1$, except if by differentiating $\bm{\E^m}$ once more with respect to the second derivatives of the form, one ends up with a vanishing expression. Moreover, this can be further used to build explicitly nontrivial Galileon $p$-form theories, by noticing that such a theory, provided it has fields equations polynomial in second derivatives of the $p$-form, should correspond to a tensor $\bm{\E^m}$ with the required symmetry, but only built out of the metric and the the Levi-Civita tensor, where $m$ is now the power of the $p$-form appearing in the Lagrangian; i.e.\ such that $\bm{\E^m}$ no longer depends on the form itself. We will exploit this remark in the following section.
\label{secIII}

\section{Construction of actions} \label{sec:construction}

\noindent
In this section we provide the explicit construction of Galileon theories, whenever they exist, corresponding to the allowed irreducible spaces listed in table \ref{tab:youngtable}. Most of the results are known (or can be inferred) from the works \cite{Deffayet:2010zh} (for cases with even $p$) and \cite{Deffayet:2016von} (for the case $p=3$), although we also find a cubic $p=4$ theory that is new to the best of our knowledge. We also emphasize, once again, that our method can be applied to construct actions for any $p$ and of any order $m$ in interactions of the field---provided they satisfy the criteria of the previous section---so that our list is only limited because of the assumption that $D\leq11$.

\subsection{Symmetric diagram theories} \label{subsec:symtableaux}

We begin by considering the construction of theories corresponding to the symmetric diagrams shown in boldface in table \ref{tab:youngtable}, i.e.\ diagrams denoted there by $\bm{(a,a)^t}$. The simplest starting point is to assume that the tensor $\bm{\E^m}$ depends only on the metric, i.e.\ it is a tensor product of the form $\eta\otimes\cdots\otimes\eta$ (see above as well as Ref.~\cite{Deffayet:2016von}). Applying the Young symmetrizer ${\bm y}^{\rm anti}_{(a,a)^t}$ on the product of metric tensors produces
\be
\bm{\epsilon_2}\equiv\bm{\epsilon}\otimes\bm{\epsilon}\,,
\ee
where $\bm{\epsilon}$ is the Levi-Civita tensor (see again \cite{Deffayet:2016von} for more details and for a review of Young symmetrizers in particular). For simplicity we will take the dimension to have its minimal value, $D=a$, where as before $a$ is the length of the columns in the diagram $(a,a)^t$. As we remarked in section \ref{sec:3formexample}, the extension to higher dimensions is achieved simply by contracting the $D-a$ additional indices in the $\bm{\epsilon}$ tensors. The tensor $\bm{\epsilon_2}$ clearly belongs to the plethysm $\sym^2(\bigwedge^D)$, while $\bm{\E^m}$ should belong to $\sym^m(\bigwedge^p) \otimes \sym^{m-1}(\sym^2)$. The relation between the two tensors can be achieved through a suitable projection of $\bm{\epsilon_2}$ onto the latter plethysm, as explained in detail in \cite{Deffayet:2016von}, but here we will take a simpler route as we are only interested in constructing explicit actions.

Recall that $\bm{\E^m}$ is obtained by differentiating $m-1$ times the equation of motion $\bm{\E}\equiv\bm{\E^1}$ w.r.t.\ $\A_{a[p],bc}$. Notice, using the so-called Littlewood--Richardson rule as well as the ``Schwarz theorem", that the latter splits under the action of the symmetric group into a direct sum of two components
\[ \ytableausetup 
{mathmode, boxsize=0.5cm, centertableaux}
\A_{a[p],bc}\quad\to\quad\begin{ytableau}
a_1 & b & c\\
a_2\\
\vdots\\
a_p\\
\end{ytableau}\quad \oplus \quad\begin{ytableau}
b & c\\
a_1\\
a_2\\
\vdots\\
a_p\\
\end{ytableau} \]
But $\bm{\E}$ must have the symmetry of a two-column Young tableau, and so it can only depend on the components of the second type, i.e.\ it can only depend on the form field through derivatives of the field strength, $\A_{[a[p],b]c}\propto\partial_c\F_{a[p]b}$, which is a known result that has been derived via other methods \cite{Henneaux:1997ha,Henneaux:1999ma}. Indeed, if $\bm{\E}$, which is polynomial in second derivatives, contained any factor of the first kind (i.e.\ one with three columns), an immediate application of the Littlewood--Richardson rule for tensor product would show that  $\bm{\E}$ should have at least three columns, which is not allowed. The bottomline is that we can obtain $\bm{\E}$ by contracting $\bm{\E^m}$ with $m-1$ powers of $\partial_c\F_{a[p]b}$ in a suitable way. The simpler route we take is to directly use $\bm{\epsilon_2}$ instead of $\bm{\E^m}$ and perform the contraction with the gauge field in such a manner that the result has the correct symmetries, which for $m=1$ are just those of $\bigwedge^p$.

\subsubsection{Maxwell-like theories}

Going back to our table \ref{tab:youngtable}, consider first the tensors $\bm{\E^2}$, which are allowed for any value of $p$, associated to the diagrams $\bm{(p+1,p+1)^t}$. Assuming that $\bm{\E^2}$ does not depend on the $p$-form, we can contract $\bm{\epsilon_2}$ with $\partial_a\F_B$ to obtain
\be
(\bm{\E}_{p,2})^{a_1\cdots a_p}=\epsilon^{a_1\cdots a_{p+1}}\epsilon^{b_1\cdots b_{p+1}}\partial_{a_{p+1}}\F_{b_1\cdots b_{p+1}}\propto \partial_b\F^{ba_1\cdots a_p}\,,
\ee
which is the linear Maxwell-like equation for a gauge $p$-form (and we use the same convention as above to denote the field equations associated to a given theory, namely $\bm{\E}_{p,m}$ designates the field equation operator of a $p$-form theory with an action $S_{p,m}$ which contains $m$ power of the field, so that the field equation contains $m-1$ such powers). It derives from the familiar action
\be
S_{p,2}=-\frac{1}{2(p+1)}\int d^Dx\,\F^{a_1\cdots a_{p+1}}\F_{a_1\cdots a_{p+1}}\,.
\ee
We focus next on the nonlinear Galileon models, first the ones with even $p$ and then the $p=3$ case.

\subsubsection{Nonlinear even $p$ Galileons}

We start by looking at the allowed nonlinear theories with even $p$, namely the ones corresponding to the diagrams $\bm{(7,7)^t}$ and $\bm{(11,11)^t}$ for $p=2$, and the diagrams $\bm{(8,8)^t}$ and $\bm{(11,11)^t}$ for $p=4$. The equation of motion of the $p=2$, $m=4$, $\bm{(7,7)^t}$ model is obtained by contracting $\bm{\epsilon_2}$ with three powers of the field strength derivative:
\be \label{eq:E24}
(\bm{\E}_{2,4})^{a_1a_2}=\epsilon^{a_1\cdots a_7}\epsilon^{b_1\cdots b_7}\partial_{a_3}\F_{b_1b_2b_3}\partial_{a_4}\F_{b_4b_5b_6}\partial_{b_7}\F_{a_5a_6a_7}\,.
\ee
The corresponding action is the one we gave in the introduction:
\be \label{eq:quartic2formaction}
S_{2,4}=\frac{1}{12}\int d^7x\,\epsilon^{a_1\cdots a_7}\epsilon^{b_1\cdots b_7}\F_{a_1a_2a_3}\F_{b_1b_2b_3}\partial_{a_4}\F_{b_4b_5b_6}\partial_{b_7}\F_{a_5a_6a_7}\,.
\ee
The $p=4$, $m=4$, $\bm{(11,11)^t}$ theory is constructed in an analogous manner. The case with $p=2$, $m=6$, $\bm{(11,11)^t}$ is also similarly built, but now the equation of motion involves five powers of the field:
\be \label{eq:E26}
(\bm{\E}_{2,6})^{a_1a_2}=\epsilon^{a_1\cdots a_{11}}\epsilon^{b_1\cdots b_{11}}\partial_{a_3}\F_{b_1b_2b_3}\partial_{a_4}\F_{b_4b_5b_6}\partial_{a_5}\F_{b_7b_8b_9}\partial_{b_{10}}\F_{a_6a_7a_8}\partial_{b_{11}}\F_{a_9a_{10}a_{11}}\,,
\ee
and follows from the action
\be
S_{2,6}=\frac{1}{18}\int d^{11}x\,\epsilon^{a_1\cdots a_{11}}\epsilon^{b_1\cdots b_{11}}\F_{a_1a_2a_3}\F_{b_1b_2b_3}\partial_{a_4}\F_{b_4b_5b_6}\partial_{a_5}\F_{b_7b_8b_9}\partial_{b_{10}}\F_{a_6a_7a_8}\partial_{b_{11}}\F_{a_9a_{10}a_{11}}\,.
\ee
So far these are the well known results first established in \cite{Deffayet:2010zh}. What we want to emphasize here is that these theories are {\it unique and nontrivial}, i.e.\ that for given $p$ and $m$ the above actions have non trivial field equations but also each provide the unique Galileon $p$-form theory with the given power $m$ of the field in its action. This follows from the group theoretic arguments outlined in section \ref{sec:classification}: the multiplicity of the irreducible corresponding to each diagram in table \ref{tab:youngtable} is equal to 1, and this corresponds to an upper bound on the number of possible inequivalent Galileon actions.\footnote{The nontrivial character of the theories follows here from a discussion similar to the one carried out in detail in Ref.\ \cite{Deffayet:2016von} for the case of the 3-form theory studied there.}

This uniqueness property is also easy to check explicitly on a case by case basis. Consider for instance the following quartic $2$-form action
\be \label{eq:quartic2formactionbis}
S_{2,4}'=\frac{3}{4}\int d^7x\,\epsilon^{a_1\cdots a_7}\epsilon^{b_1\cdots b_7}\partial_{a_1}\A_{a_2b_3}\partial_{b_1}\A_{b_2a_3}\partial_{a_4}\F_{b_4b_5b_6}\partial_{b_7}\F_{a_5a_6a_7}\,.
\ee
This theory is gauge invariant, leads to second-order field equations, and is superficially different from (\ref{eq:quartic2formaction}); however $S_{2,4}$ and $S_{2,4}'$ are in fact equal up to a numerical factor. This can be proved by using the identity
\be \label{eq:youngcondition}
\epsilon^{a_1\cdots a_nc_{n+1}\cdots c_D}\epsilon^{b_1\cdots b_n}_{\phantom{b_1\cdots b_n}c_{n+1}\cdots c_D}=\sum_{k=1}^n\epsilon^{a_1\cdots a_{n-1}b_kc_{n+1}\cdots c_D}\epsilon^{b_1\cdots a_n^{(k)}\cdots b_n}_{\phantom{b_1\cdots a_n^{(k)}\cdots b_n}c_{n+1}\cdots c_D}\,,
\ee
where the notation $a_n^{(k)}$ means that the index $a_n$ occupies the $k$th position in the list. This identity is an instance of the so-called ``Young condition'' (see e.g.\ \cite{Bekaert:2006py}): given any tensor with the index symmetries of some Young tableau $\lambda$ (in the antisymmetric basis), antisymmetrization of all indices in a column of $\lambda$ with any other index yields zero. Eq.\ (\ref{eq:youngcondition}) then follows by applying this to
\[ \ytableausetup 
{mathmode, boxsize=0.5cm, centertableaux}
\epsilon^{a_1\cdots a_nc_{n+1}\cdots c_D}\epsilon^{b_1\cdots b_n}_{\phantom{b_1\cdots b_n}c_{n+1}\cdots c_D}\quad\to\quad\begin{ytableau}
a_1 & b_1\\
a_2 & b_2\\
\vdots & \vdots\\
a_n & b_n\\
\end{ytableau} \]
We can now use this identity to relate $S_{p,4}'$ and $S_{p,4}$ for any even $p$,\footnote{The relation below is also true for odd $p$, but it is trivial in that case.} where $S_{p,4}'$ is defined as in (\ref{eq:quartic2formactionbis}) with a ``mixed'' index $\sim\partial_a\A_{a\cdots ab}$,
\be
S_{p,4}'=\frac{(p+1)}{4}\int d^Dx\,\epsilon^{a_1\cdots a_D}\epsilon^{b_1\cdots b_D}\partial_{a_1}\A_{a[p-1]b_2}\partial_{b_1}\A_{b[p-1]a_2}\partial_{a_3}\F_{b[p+1]}\partial_{b_3}\F_{a[p+1]}\,,
\ee
while $S_{p,4}$ is the ``standard'' quartic action written in terms of the field strength, as in (\ref{eq:quartic2formaction}),
\be
\begin{split}
S_{p,4}&=\frac{1}{4(p+1)}\int d^Dx\,\epsilon^{a_1\cdots a_D}\epsilon^{b_1\cdots b_D}\F_{a[p+1]}\F_{b[p+1]}\partial_{a_2}\F_{b[p+1]}\partial_{b_3}\F_{a[p+1]}\\
&=\frac{(p+1)}{4}\int d^Dx\,\epsilon^{a_1\cdots a_D}\epsilon^{b_1\cdots b_D}\partial_{a_1}\A_{a[p]}\partial_{b_1}\A_{b[p]}\partial_{a_3}\F_{b[p+1]}\partial_{b_3}\F_{a[p+1]}\,.
\end{split}
\ee
Applying (\ref{eq:youngcondition}) to $S_{p,4}'$ and performing some relabeling and reshuffling of indices, we find that
\be
S_{p,4}'=\frac{1}{2p}\,S_{p,4}\,.
\ee
We expect similar relations to hold for arbitrary values of $m$ and possibly other kinds of mixing of indices in the action.

Finally we turn our attention to the $p=4$, $m=3$, $\bm{(8,8)^t}$ irreducible space. The basic prescription we used above for finding the equation of motion ${\bm\E}$ now requires a different contraction between $\bm{\epsilon_2}$ and the derivative of the field strength:
\be \label{eq:E43}
(\bm{\E}_{4,3})^{a_1a_2a_3a_4}=\epsilon^{a_1\cdots a_8}\epsilon^{b_1\cdots b_8}\partial_{a_5}\F_{b_1b_2b_3b_4b_5}\partial_{a_6}\F_{a_7a_8b_6b_7b_8}\,.
\ee
Note that the more ``symmetric'' contraction
\be
\epsilon^{a_1\cdots a_8}\epsilon^{b_1\cdots b_8}\partial_{a_3}\F_{b_4b_5b_6b_7b_8}\partial_{b_3}\F_{a_4a_5a_6a_7a_8}\,,
\ee
may also be used (it is proportional to the one above as we will see below), but we deem the structure in (\ref{eq:E43}) more convenient for inferring the action. Indeed, it is obvious that the expression in (\ref{eq:E43}) will appear upon varying the first two terms in the action
\be \label{eq:S43}
S_{4,3}=-\frac{1}{15}\int d^{8}x\,\epsilon^{a_1\cdots a_8}\epsilon^{b_1\cdots b_8}\F_{a_1a_2a_3a_4a_5}\F_{b_1b_2b_3b_4b_5}\partial_{a_6}\F_{a_7a_8b_6b_7b_8}\,.
\ee
Less obvious is that the same is obtained from the third factor. Explicitly we find
\be
\begin{aligned}
\delta S_{4,3}&=-5\,\delta\int d^{8}x\,\epsilon^{a_1\cdots a_8}\epsilon^{b_1\cdots b_8}\partial_{a_1}\A_{a_2a_3a_4a_5}\partial_{b_1}\A_{b_2b_3b_4b_5}\partial_{a_6}\partial_{b_6}\A_{a_7a_8b_7b_8}\\
&=-5\int d^{8}x\,\epsilon^{a_1\cdots a_8}\epsilon^{b_1\cdots b_8}\bigg[-2\delta \A_{a_2a_3a_4a_5}\partial_{a_1}\partial_{b_1}\A_{b_2b_3b_4b_5}\partial_{a_6}\partial_{b_6}\A_{a_7a_8b_7b_8}\\
&\quad+\delta \A_{a_7a_8b_7b_8}\partial_{b_6}\partial_{a_1}\A_{a_2a_3a_4a_5}\partial_{a_6}\partial_{b_1}\A_{b_2b_3b_4b_5}\bigg]\,.
\end{aligned}
\ee
But two applications of the Young condition (\ref{eq:youngcondition}) show that
\be
\begin{aligned}
&\epsilon^{a_1\cdots a_9}\epsilon^{b_1\cdots b_9}\delta A_{a_7a_8b_7b_8}\partial_{b_6}\partial_{a_1}A_{a_2a_3a_4a_5}\partial_{a_6}\partial_{b_1}A_{b_2b_3b_4b_5}\\
&=-\epsilon^{a_1\cdots a_9}\epsilon^{b_1\cdots b_9}\delta A_{a_2a_3a_4a_5}\partial_{a_1}\partial_{b_1}A_{b_2b_3b_4b_5}\partial_{a_6}\partial_{b_6}A_{a_7a_8b_7b_8}\,,
\end{aligned}
\ee
which confirms our claim that (\ref{eq:E43}) derives from the action (\ref{eq:S43}).

We stress here that the 4-form theory we have just constructed, and defined e.g.\ by the action (\ref{eq:S43}), is new and has never been discussed before as far as we know. In particular, it is qualitatively different from all the known $p$-forms Galileon theories so far in that all these theories have actions containing even powers of the field, which is due to the symmetric way in which indices are contracted with the the $\bm{\epsilon}$ tensors, while our new theory (\ref{eq:S43}) on the other hand has a cubic action.

\subsubsection{Nonlinear $p=3$ Galileon}

We can construct the quartic $3$-form gauge theory that we reviewed in section \ref{sec:3formexample} in a similar way. From table \ref{tab:youngtable} we infer that there exists a tensor $\bm{\E^4}$ in the plethysm $\sym^4(\bigwedge^3) \otimes \sym^{3}(\sym^2)$ depending only on the metric; this was found explicitly in \cite{Deffayet:2016von}, but again we follow the approach of writing the equation of motion $\bm{\E}$ directly from $\bm{\epsilon_2}$, the direct product of two Levi-Civita tensors, and contract with derivatives of the field strength. An important difference between the even and odd $p$ cases is that the simplest contraction,
\be
\epsilon^{a_1\cdots a_9}\epsilon^{b_1\cdots b_9}\partial_{a_4}\F_{b_1\cdots b_4}\partial_{a_5}\F_{b_5\cdots b_8}\partial_{b_9}\F_{a_6\cdots a_9}\,,
\ee
is now identically zero. However, the alternative contraction
\be \label{eq:eom3form}
(\bm{\E}_{3,4})^{a_1a_2a_3}=\epsilon^{a_1\cdots a_9}\epsilon^{b_1\cdots b_9}\partial_{a_4}\F_{b_2b_3b_4b_5}\partial_{a_6}\F_{a_7a_8b_1b_9}\partial_{b_6}\F_{b_7b_8a_5a_9}\,
\ee
is non trivial. This is indeed the correct equation of motion of the $p=3$ theory found in \cite{Deffayet:2016von}, but written in manifestly gauge invariant form. We prove this by a direct calculation of the field equation that derives from the action
\be
S_{3,4}=\frac{1}{4}\int d^9x\,\epsilon^{a_1\cdots a_9}\epsilon^{b_1\cdots b_9}\partial_{a_1}\F_{b_2b_3b_4b_5}\partial_{b_1}\F_{a_2a_3a_4a_5}\partial_{b_6}\A_{b_7b_8a_9}\partial_{a_6}\A_{a_7a_8b_9}\,.
\ee
Varying with respect to the gauge field yields
\be
\begin{aligned}
\delta S_{3,4}&=8\int d^9x\,\epsilon^{a_1\cdots a_9}\epsilon^{b_1\cdots b_9}\bigg[\delta \A_{b_3b_4b_5}\partial_{b_1}\partial_{a_2}\A_{a_3a_4a_5}\partial_{a_1}\partial_{b_6}\A_{b_7b_8a_9}\partial_{b_2}\partial_{a_6}\A_{a_7a_8b_9}\\
&\quad-\delta \A_{b_7b_8a_9}\partial_{a_1}\partial_{b_2}\A_{b_3b_4b_5}\partial_{b_1}\partial_{a_2}\A_{a_3a_4a_5}\partial_{b_6}\partial_{a_6}\A_{a_7a_8b_9}\bigg]\,.
\end{aligned}
\ee
Using the Young condition (\ref{eq:youngcondition}) we find that the second term is in fact equal to the first, so that
\be
(\bm{\E}_{3,4})^{b_3b_4b_5}=16\,\epsilon^{a_1\cdots a_9}\epsilon^{b_1\cdots b_9}\partial_{b_1}\partial_{a_2}\A_{a_3a_4a_5}\partial_{a_1}\partial_{b_6}\A_{b_7b_8a_9}\partial_{b_2}\partial_{a_6}\A_{a_7a_8b_9}\,.
\ee
After relabeling some indices it is easy to see that this indeed matches eq.\ (\ref{eq:eom3form}). One may ask whether an independent tensor $\bm{\E}$ could be obtained in (\ref{eq:eom3form}) by performing a different contraction of indices. Our uniqueness argument again ensures that there cannot be other equations of motion that are derivable from an action, but it would be interesting to see if there may exist other theories that do not satisfy this criterion.

\subsection{Asymmetric diagram theories: redundant cases}

Next we study the asymmetric diagrams listed in table \ref{tab:youngtable}, i.e.\ the diagrams $(b,a)^t$ with $b>a$. The two important questions we want to address are (1) whether a nontrivial Galileon theory exists for a given tableau, and if so (2) whether it is independent of the actions we constructed in subsection \ref{subsec:symtableaux} for the symmetric diagrams. In this and the next subsections we show that the answer to (1) is affirmative for some, but not all, of the diagrams; however, we also show that the answer to (2) is negative, that is, the candidate asymmetric diagram theories are necessarily redundant and correspond, in a way that we will explain, to already constructed theories and therefore the models we classified above exhaust all possibilities.

Here we start by indicating which of the asymmetric diagrams of table \ref{tab:youngtable} are ``redundant'' in that they correspond to the theories already constructed in the previous subsection; these are the diagrams shown in boldface in the table. To understand why they are redundant we consider the concrete example of the $p=2$, $m=4$, $(\bm{7,7})^t$ model before we move on to the general argument. The field equation is given in (\ref{eq:E24}), which we rewrite more concisely as
\be \label{eq:E24bis}
(\bm{\E}^1)^{a_1a_2}=\epsilon^{a_1 \cdots a_7} \epsilon^{b_1 \cdots b_7}(\partial_{b_1}{\F}_{A_1})(\partial_{a_6}{\F}_{B_1})(\partial_{a_7}{\F}_{B_2})\,,
\ee
where $A_1 = \{ a_3, a_4, a_5\}$, $B_1 = \{b_2, b_3, b_4\}$, and $B_2 = \{ b_5, b_6, b_7\}$. Recall that $\bm{\E}^m$ is defined as the $(m-1)$th derivative of $\bm{\E}^1$ w.r.t.\ the second derivative of the form field. As explained in section \ref{sec:classification}, for each $m$ the tensor $\bm{\E}^m$ respects all the symmetry criteria for it to correspond to a possible Galileon theory and, moreover, it must be a linear combination of tensors with the index symmetries of two-column Young diagrams. For $D\leq11$ these diagrams are included in table \ref{tab:youngtable}. Now for a given theory, such as the one having the field equations operator given by (\ref{eq:E24bis}), this implies that the successive derivatives of the field equation w.r.t.\ the second derivative of the form field must also correspond to some diagram in this table. Explicitly, for the example in (\ref{eq:E24bis}) one gets
\be
\begin{split}
(\bm{\E}^2)^{a_1a_2[c_1c_2d_1]d_2}&=\frac{\partial(\bm{\E}^1)^{a_1a_2}}{\partial\F_{c_1c_2d_1,d_2}}\\
&=2\,\epsilon^{a_1a_2A_1d_2a_7}\epsilon^{b_1c_1c_2d_1B_2}(\partial_{b_1}{\F}_{A_1})(\partial_{a_7}{\F}_{B_2})\\
&\quad+\epsilon^{a_1a_2c_1c_2d_1a_6a_7}\epsilon^{d_2B_1B_2}(\partial_{a_6}{\F}_{B_1})(\partial_{a_7}{\F}_{B_2})\,,
\end{split}
\ee
and note that here we are not keeping track of irrelevant overall factors. For convenience we have antisymmetrized some of the indices in $\bm{\E}^2$ so that the derivative involves directly the field strength. We observe that the two terms after the last equality have index symmetries corresponding, respectively, to the Young tableaux
\[ \ytableausetup 
{mathmode, boxsize=0.5cm, centertableaux}
\begin{ytableau}
a_1 & c_1\\
a_2 & c_2\\
d_2 & d_1\\
\end{ytableau}\qquad\mbox{and}\qquad
\begin{ytableau}
a_1 & d_2\\
a_2\\
c_1\\
c_2\\
d_1\\
\end{ytableau}
\]
which precisely match the diagrams $(\bm{3,3})^t$ and $(\bm{5,1})^t$ indicated in the $m=2$, $p=2$ cell in table \ref{tab:youngtable}. Differentiating again gives
\be
\begin{split}
(\bm{\E}^3)^{a_1a_2[c_1c_2d_1]d_2[c_3c_4d_3]d_4}&=\frac{\partial(\bm{\E}^2)^{a_1a_2[c_1c_2d_1]d_2}}{\partial\F_{c_3c_4d_3,d_4}}\\
&=2\left(\epsilon^{a_1a_2c_1c_2d_1d_4a_7}\epsilon^{d_2c_3c_4d_3B_2}+\{c_{1,2},d_{1,2}\}\leftrightarrow\{c_{3,4},d_{3,4}\}\right)\partial_{a_7}{\F}_{B_2}\\
&\quad+2\epsilon^{a_1a_2A_1d_2d_4}\epsilon^{b_1c_1c_2d_1c_3c_4d_3}\partial_{b_1}{\F}_{A_1}\,,
\end{split}
\ee
and now each term entering in $\bm{\E}^3$ has the symmetries of the diagram $(\bm{6,4})^t$, which is therefore also redundant in the sense we explained before. One last differentiation yields a combination of $\bm{\epsilon_2}=\bm{\epsilon}\otimes\bm{\epsilon}$ tensors, as we knew from our bottom-up construction of the previous subsection.

We can easily generalize this argument to check which of the diagrams listed in table \ref{tab:youngtable} are related, via differentiation of the tensors $\bm{\E}^m$, to an already known Galileon vertex. Consider one such diagram $(b,a)^t$; according to the Littlewood--Richardson rule, differentiating w.r.t.\ the field strength has the effect of adding $p+1$ boxes to the first column and one box to the second, or vice versa, so that if we repeat this process $n$ times we will end up with a linear combination of tensors of symmetry types given by
\be
\big(b+j(p+1)+k\,,\,a+j+k(p+1)\big)^t\,,
\ee
where $j$ and $k$ are positive integers such that $j+k=n$. The process of course ends when all the fields have been stripped out and one is left with a combination of $\bm{\epsilon_2}$ tensors. The latter correspond to a symmetric diagram, and therefore there must exist $j$ and $k$ such that
\be
b+j(p+1)+k=a+j+k(p+1)\qquad \Rightarrow\qquad k-j=\frac{b-a}{p}\,. \label{eqn:redundant}
\ee
From this we conclude that the diagram $(b,a)^t$ will be related to a known theory if the difference $b-a$ is an integer multiple of $p$. All the asymmetric diagrams shown in boldface in table \ref{tab:youngtable} satisfy this criterion. Note that there is one particular case, namely $(\bm{11,3})^t$, which agrees with this and yet is not related to the Galileon models we classified above. The reason is simply that there we restricted our attention to $D\leq11$, while this diagram arises as a term in the tensor $\bm{\E}^4$ of an eighth-order $p=2$ vertex with minimal dimension $D=15$, which we didn't consider explicitly.

\subsection{Asymmetric diagram theories: a no-go result}

We have argued that some of the asymmetric diagrams in table \ref{tab:youngtable}, those in boldface font, are related to tensors $\bm{\E}^m$ that correspond to Galileon theories discovered/rediscovered in subsection \ref{subsec:symtableaux}. However, recall that these diagrams are only unique for a given order in the interactions of the gauge field; for instance, the $p=2$, $(\bm{7,7})^t$ diagram relates to a tensor $\bm{\E}^4$ with no fields (which was in fact our starting point in constructing the $p=2$ quartic vertex) and also to a tensor $\bm{\E}^4$ with two powers of the field that arises from differentiating the equation of motion of the $p=2$ sixth-order theory. Thus the fact that they are ``redundant'' in the sense explained above doesn't necessarily mean that they cannot be related to other theories. Furthermore, we still haven't classified the remaining asymmetric diagrams, those shown in plain font in table \ref{tab:youngtable}. The purpose of this subsection is to rule out the possibility that any of the asymmetric diagrams could correspond to an unknown theory.

Recollect from our bottom-up construction of subsection \ref{subsec:symtableaux} that, given any nontrivial equation of motion that depends only on the second derivative of the field, differentiation w.r.t.\ to the second derivative of the gauge field will eventually produce a tensor $\bm{\E}^m$ that doesn't involve the field. We assumed there that such an $\bm{\E}^m$ depended only on the metric tensor $\bm{\eta}$, which in turn implied that $\bm{\E}^m$ had the symmetry of a Young diagram with two columns of equal length (as it was proved in \cite{Deffayet:2016von}). But one may ask whether a field-independent $\bm{\E}^m$ could involve the Levi-Civita tensor $\bm{\epsilon}$ in addition to $\bm{\eta}$. That this is impossible can be easily proved as follows. Any even number of $\bm{\epsilon}$'s can be reduced to a linear combination of tensor products of metrics, and so we may assume that $\bm{\E}^m$ contains a single Levi-Civita tensor and consider tensors of the form $\epsilon \otimes \eta\otimes\cdots\otimes\eta$. In the process of decomposing such a tensor into tensors with irreducible symmetries, it is enough for us to consider the first two factors $\epsilon \otimes \eta$ of the tensor product. Indeed, these factors correspond to the Young diagrams 
\[ \ytableausetup 
{mathmode, boxsize=0.5cm, centertableaux}
\epsilon\otimes\eta\quad\to\quad
\begin{ytableau}
$$\\
$$\\
$$\\
$\vdots$\\
$$\\
\end{ytableau}
\quad\otimes\quad\begin{ytableau} $$ & $$\\
\end{ytableau} \]
which clearly doesn't contain any two-column diagram with at most $D$ boxes in each column, but only one with three columns. Hence, the full tensor product cannot contain diagrams with less than three columns. But since the allowed tensors $\bm{\E^m}$ should have the index symmetries of two-column diagrams, we conclude that any such nontrivial tensor that is independent of the $p$-form field must necessarily be a tensor product of metrics.

In summary, we have established that all nontrivial Galileon $p$-form theories must be related, through differentiation of the field equation $\bm{\E}$, to tensors $\bm{\E^m}$ which do not involve the field and hence correspond to the symmetric diagrams of table \ref{tab:youngtable}. Here, as already stated earlier, by Galileon $p$-form theories we mean that their equations of motion depend only on the second derivative of the $p$-form. The asymmetric diagrams are therefore either ``redundant'' (in boldface) if they relate to an already known $\bm{\E^m}$, or ``empty'' (in plain font) if they simply bear no connection to an existing theory. This completes our classification of the irreducible spaces that were admissible on symmetry grounds as candidates for the tensors $\bm{\E^m}$ of a Galileon $p$-form action.

Asymmetric diagrams in plain font in table \ref{tab:youngtable}, i.e.\ those that have different numbers of boxes in the two columns and that do not satisfy the condition (\ref{eqn:redundant}), do not lead to Galileon $p$-form theories but may lead to other kinds of $p$-form theories in which the equation of motion depends not only on the second derivative of the field but also on the first derivative. Indeed, the product of the field strength with either the metric or the Levi-Civita tensor generically contains a nontrivial two-column diagram. Therefore, some or all of the asymmetric diagrams in plain font in table \ref{tab:youngtable} might be related to some equation of motion that depends on not only the second derivative of the $p$-form but also the first derivative. It is worthwhile showing this explicitly as a future work.

\section{Final remarks} \label{sec:discussion}

\noindent
The goal of this paper was to classify the $p$-form Galileon gauge field theories that exist for dimensions $D\leq11$. We emphasize again that this choice was made mostly for concreteness, since the method we employed to construct the actions can be used for any $D$. Indeed, it is straightforward to enlarge our table \ref{tab:youngtable} listing the irreducible spaces for which a nontrivial tensor $\bm{\E^m}$ may exist. The results of section \ref{sec:construction} imply that only the symmetric diagrams $(a,a)^t$ can correspond to independent Galileon theories, while asymmetric ones may be ignored as far as one is only concerned with identifying nontrivial vertices. For instance there exist fourth- and sixth-order vertices for a Galileon $5$-form,\footnote{Note that because of the inequality (\ref{eq:mupperbound}) we know that there cannot be any Galileon $3$-form actions beyond the one we have presented.} schematically
\be
S_{5,4}\sim\int d^{13}x\,\epsilon^{a\cdots}\epsilon^{b\cdots}\partial_{a}\partial_{b}\A_{b[5]}\partial_{b}\partial_{a}\A_{a[5]}\partial_{a}\A_{a[4]b}\partial_{b}\A_{b[4]a}\,,
\ee
and
\be
S_{5,6}\sim\int d^{20}x\,\epsilon^{a\cdots}\epsilon^{b\cdots}\partial_{a}\partial_{b}\A_{b[5]}\partial_{b}\partial_{a}\A_{a[5]}\partial_{a}\partial_{b}\A_{a[4]b}\partial_{b}\partial_{a}\A_{b[4]a}\partial_{a}\A_{a[3]b[2]}\partial_{b}\A_{b[3]a[2]}\,,
\ee
again using the minimal dimension $D$ in each case. It is straightforward to check that these actions are gauge invariant and lead to field equations of second order. To summarize our resulting classification, we have shown that for $D\leq11$, the only nontrivial Galileon $p$-form theories are either the ones obtained in \cite{Deffayet:2010zh} (for even $p$), or the $p=3$ theory found in \cite{Deffayet:2016von}, or a $p=4$ cubic theory first introduced in this work (with the action given in eq.~(\ref{eq:S43})).

There are a number of related avenues of research that can be pursued. First, the theories we have classified are Galileon terms in the strict sense, i.e.\ they have equations of motion involving only the second derivative of the field. It would be interesting to see to what extent the methods of \cite{Deffayet:2016von} can be generalized to allow for operators with one and zero derivatives. This should certainly relax some of the no-go results that affect Galileon $p$-forms; for example the $1$-form Chern--Simons action leads to a one-derivative term in the field equation. Another intriguing question is how to systematically construct Galileon interactions with more than one field, be they of the same degree $p$ (i.e.\ a set of interacting ``colored'' $p$-forms) or different types of forms. That this is possible was already shown in \cite{Deffayet:2010zh}, but a full classification of such ``mixed'' interaction vertices is lacking. Finally, the problem of how to covariantize Galileon $p$-form theories hasn't been fully solved. Minimal coupling of Galileons to a dynamical metric generically leads to higher derivative terms in equations of motion and thus possibly to ghost instabilities in the gravitational sector, which can be remedied, in the case of scalars \cite{Deffayet:2009wt} and some even $p$-forms \cite{Deffayet:2010zh}, by the addition of nonminimal couplings to the Riemann tensor.\footnote{See, however, e.g.\ refs.~\cite{Gleyzes:2014dya,Lin:2014jga,Deffayet:2015qwa,Langlois:2015cwa} for theories in which the equations of motion contain higher derivatives but extra degrees of freedom do not appear.} Whether this can be done in general is an open question that we will address in a forthcoming work.

\begin{acknowledgments}
\noindent
We are grateful to Nicolas Boulanger for some very helpful conversations. The work of C.D., S.G.-S., and V.S., as well as visits of S.M., were supported by the European Research Council under the European Community's Seventh Framework Programme (FP7/2007-2013 Grant Agreement no.\ 307934, NIRG project). The work of S.M.\ was supported by Japan Society for the Promotion of Science (JSPS) Grants-in-Aid for Scientific Research (KAKENHI) No.\ 24540256, No.\ 17H02890, No.\ 17H06359, No.\ 17H06357 and by World Premier International Research Center Initiative (WPI), MEXT, Japan.
\end{acknowledgments}

\bibliographystyle{apsrev4-1}
%\bibliography{pGalileonDraft_bibliography.bib}
\bibliography{pGalileonDraft_bibliography}

\end{document}